\documentclass[12pt,letter]{article}
\usepackage{graphicx, epsfig, color,cite}
\usepackage{amsmath}
\usepackage{amssymb}
\usepackage{subfigure}
\def\bi{\begin{itemize}}
\def\ei{\end{itemize}}

\def\tst{\tilde t}

\def\tg{\tilde g}

\def\tw{\widetilde W}
\def\tz{\widetilde Z}

\def\alt{\lesssim}
\def\agt{\gtrsim}
\def\be{\begin{equation}}  
\def\ee{\end{equation}}  
\def\bea{\begin{eqnarray}}  
\def\eea{\end{eqnarray}}

\begin{document}
\begin{titlepage}
\begin{flushright}
OU-HEP-160225
\end{flushright}

\vspace{0.5cm}
\begin{center}
{\Large \bf The Higgs mass and natural supersymmetric spectrum\\
 from the landscape
}\\ 
\vspace{1.2cm} \renewcommand{\thefootnote}{\fnsymbol{footnote}}
{\large Howard Baer$^1$\footnote[1]{Email: baer@nhn.ou.edu },
Vernon Barger$^2$\footnote[2]{Email: barger@pheno.wisc.edu},
Michael Savoy$^1$\footnote[3]{Email: savoy@nhn.ou.edu}
and Hasan Serce$^1$\footnote[4]{Email: serce@ou.edu}
}\\ 
\vspace{1.2cm} \renewcommand{\thefootnote}{\arabic{footnote}}
{\it 
$^1$Dept. of Physics and Astronomy,
University of Oklahoma, Norman, OK 73019, USA \\[3pt]
}
{\it 
$^2$Dept. of Physics,
University of Wisconsin, Madison, WI 53706 USA \\[3pt]
}

\end{center}

\vspace{0.5cm}
\begin{abstract}
\noindent
In supersymmetric models where the superpotential $\mu$ term 
is generated with  $\mu\ll m_{soft}$ 
({\it e.g.} from radiative Peccei-Quinn symmetry breaking or compactified string models with
sequestration and stabilized moduli), and where the string landscape
1. favors soft supersymmetry (SUSY) breaking terms as large as possible and 
2. where the anthropic condition that electroweak symmetry is properly broken 
with a weak scale $m_{W,Z,h}\sim 100$ GeV ({\it i.e.} not too weak of weak interactions),
then these combined landscape/anthropic requirements act as an {\it attractor} pulling 
the soft SUSY breaking terms towards values required by models with radiatively-driven naturalness: 
near the line of criticality where electroweak symmetry is barely broken
and the Higgs mass is $\sim 125$ GeV.
The pull on the soft terms serves to ameliorate the SUSY flavor and CP problems.
The resulting sparticle mass spectrum may barely be accessible at high-luminosity LHC 
while the required light higgsinos should be visible at a linear $e^+e^-$ collider with $\sqrt{s}>2m(higgsino)$.

\end{abstract}
\end{titlepage}
%\pacs{12.60.-i, 95.35.+d, 14.80.Ly, 11.30.Pb}
%12.60.-i   Models beyond the standard model
%95.35.+d   Dark matter

The Standard Model is afflicted with several naturalness problems: 
\begin{enumerate}
\item in the electroweak sector, why is the Higgs mass $m_h\simeq 125$ GeV so light when quadratic 
divergences seemingly destabilize its mass\cite{susskind} and 
\item why is the QCD Lagrangian term $\frac{\bar{\theta}}{32\pi^2}G_{A\mu\nu}\tilde{G}_A^{\mu\nu}$ so tiny 
($\bar{\theta}\alt 10^{-10}$ from measurements of the neutron electric dipole moment) when its existence 
seems a necessary consequence of the $\theta$ vacuum solution to the  $U(1)_A$ problem 
(the strong $CP$ problem)\cite{peccei}?
\item A third naturalness problem emerges when gravity is included into the picture: 
why is the cosmological constant $\Lambda\simeq 10^{-47}$ GeV$^4\ll M_P^4$ so small when there is 
no known mechanism for its suppression\cite{weinberg}?
\end{enumerate}
Each of these problems requires an exquisite fine-tuning of parameters to maintain accord with experimental data.
Such fine-tuning is thought to represent some pathology with or missing element within the underlying theory
and cries out for a ``natural'' solution in each case.

The most compelling solution to problem \#1 is to extend the spacetime symmetry structure which 
underlies quantum field theory to include its most general structure: 
the super-Poincare group which includes supersymmetry (SUSY) 
transformations\cite{witten,wss}. 
The extended symmetry implies a Fermi-Bose correspondence which guarantees cancellation of
quadratic divergences to all orders in perturbation theory. 
Supersymmetrization of the SM implies the existence of superpartner matter states with 
masses of order $M_S\sim 1$ TeV\cite{susyreviews,wss}. 
Searches are underway at the CERN LHC for evidence of the superpartner matter states.

The most compelling solution to problem \#2 is to postulate an additional spontaneously broken global 
Peccei-Quinn (PQ) symmetry and its concommitant
axion field $a$ which induces additional potential contributions that 
allow the offending $CP$ violating term to dynamically settle to a tiny value\cite{pqww,ksvz,dfsz}. 
Searches for the physical axion field are proceeding at experiments like ADMX\cite{admx} 
but so far sensitivity has barely 
reached parameter values needed to solve the strong $CP$ problem. 

At present the leading solution to problem \#3 is the hypothesis of the landscape: a vast number of string 
theory vacua states each with different physical constants\cite{landscape}. 
In this case, the cosmological constant ought to be present, 
but if it is too large, then the universe would expand too quickly to allow for 
galaxy condensation and there would be no observers present to measure $\Lambda$. 
This ``anthropic'' explanation for the magnitude of $\Lambda$ met with
great success by Weinberg\cite{weinberg2} who was able to predict its value to within a factor 
of a few even well before it was measured\cite{Lambda}.

While the SUSY solution to the scalar mass problem seems convincing at the level of quadratic divergences, 
there is a high level of concern that the fine-tuning problem has re-arisen in light of 
1. the apparently severe LHC bounds on sparticle masses and 
2. the rather high measured value of $m_h$. 
This perception arises from two viewpoints on measuring naturalness.
\begin{itemize}
\item Log-divergent contributions to the Higgs mass 
$\delta m_h^2\sim \frac{-3f_t^2}{16\pi^2}m_{\tst}^2\log\left(\Lambda^2/m_{\tst}^2\right)$ become large
for TeV-scale top squark masses $m_{\tst}$ and $\Lambda$ as high as $m_{GUT}\simeq 2\times 10^{16}$ 
GeV\cite{harnik}. 
This argument has been challenged in that a variety of inter-dependent log terms, 
some positive and some negative, contribute to the Higgs mass. 
Evaluation of the combined log terms via renormalization group equations reveals the 
possibility of large cancellations in evaluation of the Higgs mass\cite{comp3,seige}.
\item The EENZ/BG fine-tuning measure\cite{eenz} 
$\Delta_{\rm BG}=max_i|\frac{\partial\log m_Z^2}{\partial\log p_i}|$ (where $p_i$ are fundamental parameters 
of the theory) is traditionally evaluated using the various soft SUSY breaking terms as fundamental
parameters. In this case, low $\Delta_{\rm BG}$ favors sparticle masses in the 100 GeV range. 
These evaluations have been challenged in that in more fundamental theories, 
the soft terms are not independent, but are derived in terms of more fundamental quantities, 
for instance the gravitino mass $m_{3/2}$ in supergravity theories. 
Evaluation of $\Delta_{\rm BG}$ instead in terms of $\mu^2$ and $m_{3/2}^2$ 
allows for just $\mu$ and $m_{H_u}$ to be $\sim 100$ GeV while the other sparticles
can safely lie at or beyond the TeV scale\cite{comp3,seige}. 
\end{itemize}

A more conservative measure which is in accord with the above (corrected) measures is to evaluate just the 
weak scale contributions to the $Z$ mass.
The minimization condition for the Higgs potential  $V_{\rm tree} + \Delta V$ in the 
minimal supersymmetric Standard Model (MSSM) reads
\be 
\frac{m_Z^2}{2} = \frac{m_{H_d}^2 + \Sigma_d^d -
(m_{H_u}^2+\Sigma_u^u)\tan^2\beta}{\tan^2\beta -1} -\mu^2 \; .
\label{eq:mzs}
\ee 
The radiative corrections $\Sigma_u^u$ and $\Sigma_d^d$ include contributions from
various particles and sparticles with sizeable Yukawa and/or gauge
couplings to the Higgs sector.
Expressions for the $\Sigma_u^u$ and $\Sigma_d^d$ are given in the Appendix of Ref. \cite{rns}.

A naturalness measure $\Delta_{\rm EW}$ has been introduced\cite{ltr,rns} which
compares the largest contribution on the right-hand-side of Eq. \ref{eq:mzs} 
to the value of $m_Z^2/2$. If they are comparable ($\Delta_{\rm EW}\alt 10-30$), 
then no unnatural fine-tunings are required to generate $m_Z=91.2$ GeV. 
The main requirement for low fine-tuning is then that
\bi
\item $|\mu |\sim m_Z$\cite{Chan:1997bi,Barbieri:2009ew,hgsno} 
(with $\mu \agt 100$ GeV to accommodate LEP2 limits 
from chargino pair production searches) and also that 
\item $m_{H_u}^2$ is driven radiatively to small, and not large, negative values~\cite{ltr,rns}. 
Also, 
\item the top squark contributions to the radiative corrections $\Sigma_u^u(\tst_{1,2})$ 
are minimized for TeV-scale highly mixed top squarks\cite{ltr}. 
This latter condition  also lifts the Higgs mass  to $m_h\sim 125$ GeV.
\item First and second generation squark and slepton masses may range as high as 10-20 TeV with little cost to
naturalness\cite{rns,upper}.
\ei

The typical low $\Delta_{\rm EW}$ SUSY mass spectra is characterized by 
1. a set of light higgsinos $\tw_1^\pm$ and $\tz_{1,2}$ with masses $\sim 100-200$ GeV, 
2. gluinos with mass $m_{\tg}\sim 1.5-4$ TeV, 
3. highly mixed stops with mass $m_{\tst_1}\alt 3$ TeV and $m_{\tst_2}\alt 8$ TeV.
Several versions of supergravity GUT models have been found to generate such 
``natural'' spectra\cite{guts}.
For instance, the two-extra-parameter non-universal Higgs mass model\cite{nuhm2} 
(NUHM2) with matter scalars 
$m_0\sim 3-10$ TeV, $m_{1/2}\sim 0.5-2$ TeV, $A_0\sim\pm(1-2)m_0$ and $\tan\beta\sim 10-30$ 
with $m_{H_u}\sim (1.3-2)m_0$ and $m_{H_d}\sim m_A\sim 1-8$ TeV
produces spectra with $\Delta_{\rm EW}\alt 30$. 
In particular, the up-Higgs soft mass is as large as possible such that
the RG running of $m_{H_u}^2$ nearly cancels out its GUT-scale boundary value $m_{H_u}^2(\Lambda )$, {\it i.e.}
$m_{H_u}^2$ runs to small weak scale values $\sim -(100-200)^2$ GeV$^2$ so that electroweak symmetry is 
{\it barely broken}. The soft terms, especially $m_{H_u}^2$, lie on the edge of {\it criticality}: if $m_{H_u}^2$
is much bigger, then EW symmetry does not get broken while if $m_{H_u}^2(\Lambda )$ is much smaller, then
it would likely generate a value of $m_Z$ far beyond its measured value of 91.2 GeV.

While such effective theory parameters can successfully generate natural SUSY mass spectra, the question arises:
is there some mechanism which favors parameters which barely break EW symmetry, and which generate a weak scale
$m_{W,Z,h}\sim 100$ GeV rather than say in the TeV range? 
In this letter, we argue that the string landscape-- which provides
some understanding for the small but non-zero cosmological constant-- also favors soft SUSY breaking 
terms as large as possible such that they generate a universe which is habitable for observers: 
if the soft parameters were much larger, 
then they would lead to a vacuum state with color breaking minima, or unbroken EW symmetry or 
if they were much smaller they would generate a weak scale characterized by the TeV regime. 
In the latter case, with $m_{W,Z,h}\sim 1-10$ TeV, then
weak interactions would be far weaker than in our universe: then for instance nuclear fusion reactions 
would be sufficiently suppressed so that heavy element production in stars and in the early universe 
would be far different from that of our universe, 
likely leading to a universe with chemistry unsuitable for life forms as we known them.

This topic of anthropic selection of soft SUSY breaking terms has been addressed previously by
Giudice and Rattazzi\cite{GR} with some follow-up work in Ref's \cite{NP} (for mixed moduli-anomaly mediated SUSY
breaking models) and \cite{DM} (for mSUGRA/CMSSM model). 
One of the main differences of our work here is in 
the treatment of the superpotential $\mu$ parameter and the so-called SUSY $\mu$ problem. 
Under the Giudice-Masiero mechanism\cite{GM}, 
where $\mu$ arises from Higgs doublet couplings to the hidden sector via 
the Kahler potential, then $\mu$ is expected to have magnitude of order the other soft terms: 
$|\mu |\sim m_{3/2}$. 
Alternatively, in the Kim-Nilles mechanism\cite{KN}-- which is assumed here
as an axionic solution to the strong CP problem-- 
$\mu$ is initially forbidden by the requirement of Peccei-Quinn symmetry, 
but is then re-generated upon spontaneous PQ symmetry breaking at a scale $f_a\sim 10^{11}$ GeV 
with a value $\mu\sim f_a^2/M_P\ll m_{3/2}$. 
In models where PQ symmetry breaking is induced radiatively, then values of 
$m_{3/2}\sim 10$ TeV easily produce $\mu$ values around 100-200 GeV\cite{msy,radpq}.
%Since $m_h^2\sim \mu^2+m_{H_u}^2$ in the MSSM\cite{natsumi}, 
In classes of compactified string models with sequestration between the visible sector 
and the SUSY breaking sector and with stabilized moduli fields\cite{quevedo}, 
it is also found that $\mu\ll M_S$ where $M_S$ stands for the approximate scale of the collective soft
SUSY breaking terms.
In this letter we will assume 
\bi
\item the superpotential $\mu$ term has been generated by some mechanism 
such as Ref. \cite{radpq} or Ref. \cite{quevedo} to be small, comparable to $m_h= 125$ GeV. 
\ei
Then, instead of fixing $m_Z$ at its measured value, we will invert the usual
usage of Eq. \ref{eq:mzs}
to calculate $m_{W,Z,h}\sim  m_{weak}$ as an {\it output} depending on high scale values of
the soft terms and a small value of $\mu$.\footnote{In this case, 
low values of $\Delta_{\rm EW}$ can be re-interpreted as the likelihood to 
generate the weak scale $m_{weak}\sim 100$ GeV: {\it i.e.} 
$m_{weak}=\sqrt{\Delta_{\rm EW}m_Z^2/2}$.}
%Another difference is that we now know the existence of and mass value of the light Higgs scalar 
%and what this implies for soft terms:
%basically, electroweak naturalness requires a large trilinear soft term $A_0$ 
%which is automatically generated in gravity-mediated SUSY breaking models.

In the following, we will assume gravity-mediated supersymmetry breaking\cite{sugra}. 
Gravity-mediation is supported by the large value of $m_h\sim 125$ GeV 
which requires a large trilinear $A_0$ term (generic in gravity-mediation) to provide substantial 
mixing in the stop sector and consequently a boost
in the radiative corrections to the light Higgs mass\cite{mhiggs,h125}.
Gravity-mediated SUSY breaking can be parametrized by the presence of a spurion 
superfield $X=1+\theta^2 F_X$ where the auxiliary field $F_X$ obtains a vev which we
also denote by $F_X$ (here $\theta$ are anti-commuting superspace coordinates). 
Under SUSY breaking via the superHiggs mechanism, then the gravitino
gains a mass $m_{3/2}\sim F_X/M_P$ where $M_P=2.4\times 10^{18}$ GeV is the reduced 
Planck mass. The soft SUSY breaking terms are then all calculable as multiples of $m_{3/2}$\cite{sw}. 
Motivated by supergravity grand unified theories (SUSY GUTs), 
here we assume the soft breaking terms valid at $Q=m_{GUT}\simeq 2\times 10^{16}$ GeV include 
$m_0$ (a common matter scalar mass term), $m_{1/2}$ (a common gaugino mass), $A_0$ 
(a common trilinear soft term) and $B$. The latter soft term can be traded for the more
common ratio of Higgs vevs $\tan\beta\equiv v_u/v_d$ via the electroweak minimization 
conditions. We also assume separate Higgs scalar soft terms $m_{H_u}^2$ and $m_{H_d}^2$
since the Higgs superfields live in different GUT representations than matter superfields\cite{nuhm2}.
It is convenient to denote collectively the superpartner mass scale $M_S$ 
as the generic scale of soft terms.

It is reasonable to assume in the landscape that {\it any} value of 
the complex-valued field $F_X$ is equally likely. In this case, one expects the magnitude
of soft breaking terms to statistically scale linearly in $M_S$ (the likelihood of a given
value of $M_S$ is proportional to the area of an annulus $2\pi F_X\delta F_X$ in the complex
$F_X$ plane). This is important because then we see a statistical draw of soft terms towards
their largest values possible (while $\mu$ remains far smaller). 
In Ref. \cite{GR}, additional arguments are presented
that the likelihood of soft terms $M_S$ scale as a power of $M_S$; for our purposes here, 
we merely rely on a likely statistical draw by the landscape of vacua towards higher values
of soft terms. This draw is to be balanced by the anthropic requirements that
1. electroweak symmetry is appropriately broken (no charge or color breaking minima of 
the Higgs potential) and 
2. that the weak scale is typified by the values of 
$m_{weak}\sim m_{W,Z,h}\sim 100$ GeV. 
Rates for nuclear fusion reactions and beta decays all scale as $1/m_{weak}^4$ so that
heavy element production in BBN and in stars would be severely altered for too large
a value of $m_{weak}$; see Ref's \cite{kribs} for discussion.

Armed with a notion of both the statistical and anthropic pull from the landscape, we may examine the
soft SUSY breaking terms. First, we expect the matter scalar mass $m_0$ as large as possible while
maintaining $m_{weak}\sim 100$ GeV. If $m_0$ gets much beyond the 10 TeV scale, 
then the weak scale top squark masses $m_{\tst_{1,2}}$ become too large, increasing the
radiative corrections $\Sigma_u^u(\tst_{1,2})$ in Eq. \ref{eq:mzs}. For fixed $\mu\sim 100-200$ GeV,
then this increases the resultant weak scale well beyond the anthropic target 100-200 GeV.
Re-interpreting the limits on $m_0$ from Ref's \cite{rns,upper} requires $m_0\alt 10$ TeV for
$m_{weak}\sim 100$ GeV. Such large values of $m_0$ go a long ways towards solving the
SUSY flavor and CP problems via a decoupling solution\cite{dine}.

Likewise, we expect the gaugino mass $m_{1/2}$ as large as possible whilst maintaining
$m_{weak}\sim 100-200$ GeV. If the gaugino masses are too large, then they feed into the
stop masses via RG running and again the $\Sigma_u^u(\tst_{1,2})$ become too large.
For $m_{weak}\sim 100$ GeV, then typically $m_{1/2}\alt 2$ TeV leading to a gluino mass 
bound $m_{\tg}\alt 4-5$ TeV: well above the reach of LHC14\cite{andre}.
\begin{figure}[tbp]
\begin{center}
\includegraphics[height=0.3\textheight]{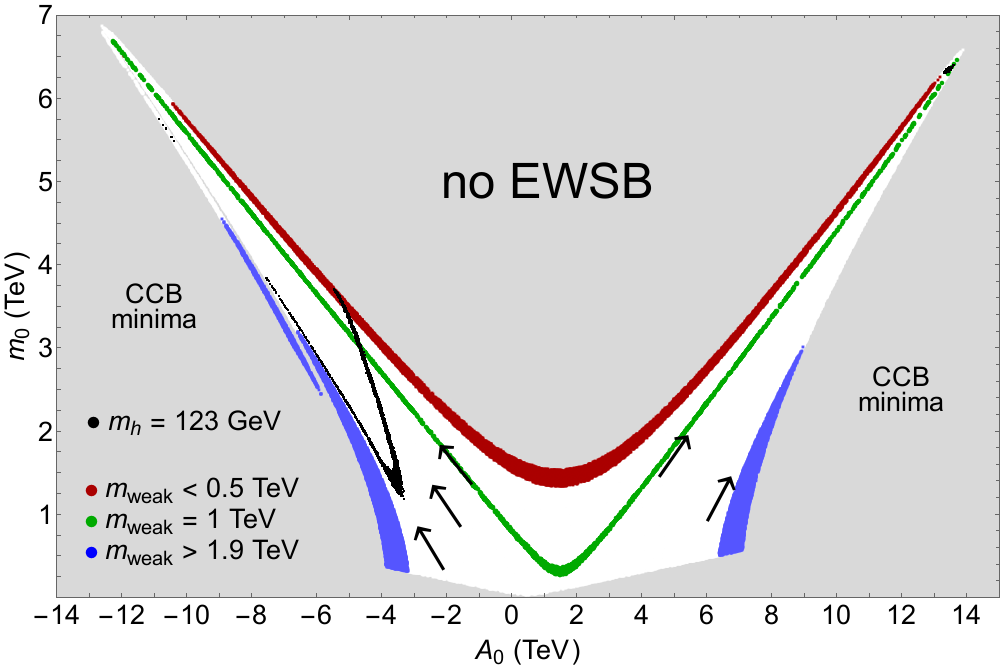}
\caption{Contours of $m_{weak}$ in the $A_0$ vs. $m_0$ plane
for $m_{1/2}=1$ TeV, $m_{H_u}=1.3 m_0$, $\tan\beta =10$ and $m_{H_d}=1$ TeV.
The arrows show the direction of statistical/anthropic pull on soft SUSY breaking terms.
\label{fig:A0_m0}}
\end{center}
\end{figure}

What of the trilinear soft term $A_0$? In Fig. \ref{fig:A0_m0} we show the 
$A_0$ vs. $m_0$ plane for the NUHM2 model with $m_{1/2}$ fixed at 1 TeV, $\tan\beta =10$
and $m_{H_d}=1$ TeV. We take $m_{H_u}=1.3 m_0$. 
The plane is qualitatively similar for different reasonable parameter choices. 
We expect $A_0$ and $m_0$ statistically to be drawn as large as possible
while also being anthropically drawn towards $m_{weak}\sim 100-200$ GeV, labelled as
the red region where $m_{weak}<500$ GeV. The blue region has $m_{weak}>1.9$ TeV and the green
contour labels $m_{weak}=1$ TeV. The arrows denote the combined 
statistical/anthropic pull on the soft terms: towards large soft terms but low $m_{weak}$.
The black contour denotes $m_h=123$ GeV with the regions to the upper left 
(or upper right, barely visible) containing larger values of $m_h$. 
We see that the combined pull on soft terms brings us to
the region where $m_h\sim 125$ GeV is generated. 
This region is characterized by highly mixed TeV-scale top squarks\cite{mhiggs,h125}. 
If instead $A_0$ is pulled too large,
then the stop soft term $m_{U_3}^2$ is driven tachyonic resulting in charge and color
breaking minima in the scalar potential (labelled CCB). 
If $m_0$ is pulled too high for fixed $A_0$, then electroweak symmetry isn't even broken.
\begin{figure}[tbp]
\begin{center}
\includegraphics[height=0.3\textheight]{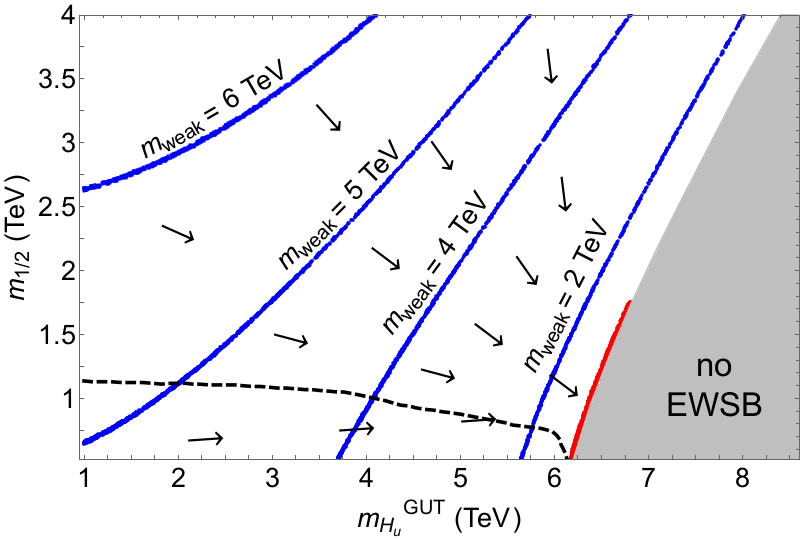}
\caption{Contours of $m_{weak}$ (blue) in the $m_{H_u}$ vs. $m_{1/2}$ plane
for $m_0=5$ TeV, $A_0=-8$ TeV, $\tan\beta =10$ and $m_{H_d}=1$ TeV.
Above the black dashed contour is where $m_h>124$ GeV.
The red region has $m_{weak}<0.5$ TeV.
The arrows show the direction of the statistical/anthropic pull on soft SUSY breaking terms.
\label{fig:mHu_mhf}}
\end{center}
\end{figure}

\begin{figure}[tbp]
\begin{center}
\includegraphics[height=0.4\textheight]{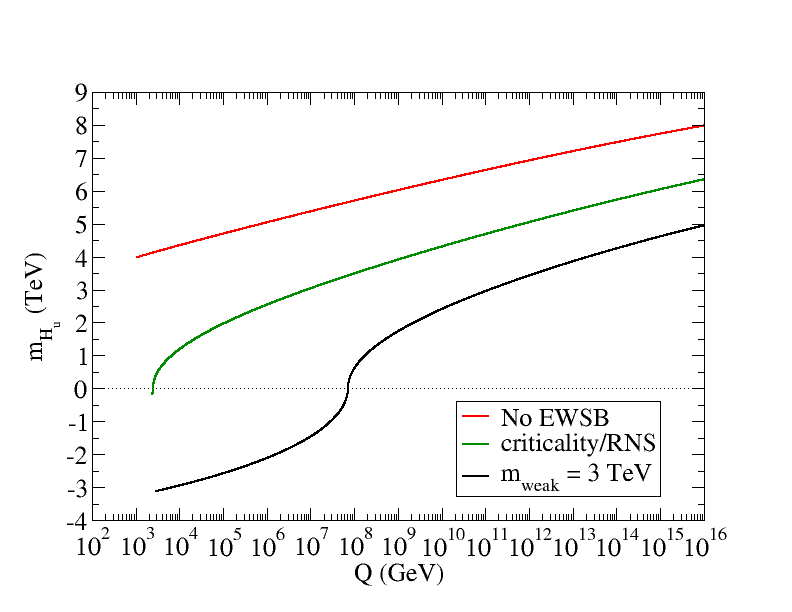}
\caption{Evolution of the soft SUSY breaking mass squared term
$sign(m_{H_u}^2)\sqrt{|m_{H_u}^2|}$ vs. $Q$ for the case of no EWSB (upper), 
criticality (middle) as in radiatively-driven natural SUSY (RNS) 
and $m_{weak}\sim 3$ TeV (lower).
Most parameters are the same as in Fig. \ref{fig:mHu_mhf}.
\label{fig:mHuQ}}
\end{center}
\end{figure}

In Fig. \ref{fig:mHu_mhf}, we show contours of $m_{weak}$ in the $m_{H_u}$ vs. $m_{1/2}$
plane for $m_0=5$ TeV, $A_0=-8$ TeV, $\tan\beta =10$ and $m_{H_d}=1$ TeV. The statistical 
flow is to large values of soft terms but the anthropic flow is towards the red region where
$m_{weak}<0.5$ TeV.
While $m_{1/2}$ is statistically drawn to large values, if it is too large then,
as before, the $\tst_{1,2}$ become too heavy and the $\Sigma_u^u(\tst_{1,2})$ become
too large so that $m_{weak}$ becomes huge.
The arrows denote the direction of the combined statistical/anthropic flow. 
The region above the black dashed contour has $m_h>124$ GeV. 
The value of $m_{H_u}(GUT)$ would like to be statistically as large as possible but if it is too large
then EW symmetry will not break. Likewise, if $m_{H_u}(GUT)$ is not large enough, then it is driven to
large negative values so that $m_{weak}\sim$ the TeV regime and weak interactions are too weak.
The situation is shown in Fig. \ref{fig:mHuQ} where we show the running of 
$sign(m_{H_u}^2)\sqrt{|m_{H_u}^2|}$ versus energy scale $Q$ 
for several values of $m_{H_u}^2(GUT)$ for $m_{1/2}=1$ TeV and with other parameters
the same as Fig. \ref{fig:mHu_mhf}. Too small a value of $m_{H_u}^2(GUT)$ leads to too large
a weak scale while too large a value results in no EWSB. The combined statistical/anthropic
pull is for barely-broken EW symmetry where soft terms teeter on the edge of criticality:
between breaking and not breaking EW symmetry. This yields the other naturalness condition
that $m_{H_u}$ is driven small negative: then the weak interactions are of the necessary strength.
These are just the same conditions for supersymmetric models with radiatively-driven 
natural SUSY (RNS)\cite{ltr,rns}.

{\it Summary:} The naturalness condition of no large unnatural cancellations in 
$m_{Z,h}$ requires small higgsino mass $\mu\sim 100-200$ GeV, $m_{H_u}^2$ driven small
rather than large negative and not-too-large radiative corrections $\Sigma_u^u (i)$.
There are mechanisms where $\mu\ll M_S$-- such as radiative PQ breaking-- but is it 
merely luck that the soft terms are poised to be just large enough to guarantee 
also that $m_{weak}\sim 100$ GeV? 
Here, we argue that the statistical landscape pull towards large soft terms coupled with 
the anthropic pull towards the Goldilocks condition-- 
small enough to break EW symmetry but not so small as to suppress weak interactions-- 
gives the required conditions for SUSY with 
radiatively-driven naturalness and barely broken EW symmetry.
While sparticles may barely be accessible to LHC, the required light higgsinos 
should be accessible to an $e^+e^-$ collider with $\sqrt{s}>2 m(higgsino)$.
We also expect ultimately detection of a higgsino-like WIMP\cite{wimp} 
along with the axion.

{\it Acknowledgements:} 
We thank Jake Baer for an inspiring essay on the landscape
and Xerxes Tata for comments on the manuscript.
This work was supported in part by the US Department of Energy, Office
of High Energy Physics.

%%%%%%%%%%%%%%%%%%%%%%%%%%%%%%%%%%%%%%%%%%%%%%%%%%%%%%

%

\begin{thebibliography}{99}
%%%%%%%%%%%%%%%%%%%%%%%%%%%%%%%%%%%%%%%%%%%%%%%%%%%%%%

%
\bibitem{susskind} L.~Susskind,
  %``Dynamics of Spontaneous Symmetry Breaking in the Weinberg-Salam Theory,''
  Phys.\ Rev.\ D {\bf 20} (1979) 2619.
%  doi:10.1103/PhysRevD.20.2619.
%
\bibitem{peccei} R.~D.~Peccei,
  %``The Strong {CP} Problem,''
  Adv.\ Ser.\ Direct.\ High Energy Phys.\  {\bf 3} (1989) 503.
%  doi:10.1142/9789814503280_0013
%
\bibitem{weinberg} S.~Weinberg,
  %``The Cosmological Constant Problem,''
  Rev.\ Mod.\ Phys.\  {\bf 61} (1989) 1.
%  doi:10.1103/RevModPhys.61.1
%
\bibitem{witten} E.~Witten,
  %``Dynamical Breaking of Supersymmetry,''
  Nucl.\ Phys.\ B {\bf 188}, 513 (1981); R.~K.~Kaul,
  %``Gauge Hierarchy in a Supersymmetric Model,''
  Phys.\ Lett.\ B {\bf 109}, 19 (1982).
%
\bibitem{wss} H.~Baer and X.~Tata,
  ``Weak scale supersymmetry: From superfields to scattering events,''
  Cambridge, UK: Univ. Pr. (2006) 537 p.
%
\bibitem{susyreviews} S.~P.~Martin,
  %``A Supersymmetry primer,''
  Adv.\ Ser.\ Direct.\ High Energy Phys.\  {\bf 21} (2010) 1.
%
\bibitem{pqww} R.~D.~Peccei and H.~R.~Quinn,
  %``CP Conservation in the Presence of Instantons,''                                                  
  Phys.\ Rev.\ Lett.\ {\bf 38}, 1440 (1977);
 S.~Weinberg, Phys.\ Rev.\ Lett.\ {\bf 40}, 223 (1978);
 F.~Wilczek, Phys.\ Rev.\ lett.\ {\bf 40}, 279 (1978).
%
\bibitem{ksvz} J.~E.~Kim, Phys.\ Rev.\ Lett.\ {\bf 43}, 103 (1979);
 M.~A.~Shifman, A.~Vainstein and V.~I.~Zakharov,
 Nucl.\ Phys.\ B {\bf 166}, 493 (1980).
%
\bibitem{dfsz} M.~Dine, W.~Fischler and M.~Srednicki,
Phys.\ Lett.\ B {\bf 104}, 199 (1981);
A.~P.~Zhitnitskii, Sov.\ J.\ Phys.\ {\bf 31}, 260 (1980).
%
\bibitem{admx} L.~Duffy, {\it et.~al.}, Phys.\ Rev.\ Lett.\ {\bf 95}, 091304 (2005)
and Phys.\ Rev.\ D {\bf 74}, 012006 (2006);
for a review, see S.~J.~Asztalos, L.~Rosenberg, K.~van Bibber, P.~Sikivie
and K.~Zioutas,
  %``Searches for astrophysical and cosmological axions,''                                             
Ann.\ Rev.\ Nucl.\ Part.\ Sci.\ {\bf 56}, 293 (2006).
%
\bibitem{landscape} R.~Bousso and J.~Polchinski,
  %``Quantization of four form fluxes and dynamical neutralization of the cosmological constant,''
  JHEP {\bf 0006} (2000) 006;
L.~Susskind,
  %``The Anthropic landscape of string theory,''
%  In *Carr, Bernard (ed.): Universe or multiverse?* 247-266
  hep-th/0302219;
M.~R.~Douglas,
  %``The Statistics of string / M theory vacua,''
  JHEP {\bf 0305} (2003) 046;
M.~Dine, E.~Gorbatov and S.~D.~Thomas,
  %``Low energy supersymmetry from the landscape,''
  JHEP {\bf 0808} (2008) 098;
for reviews, see M.~Dine,
  %``Supersymmetry, naturalness and the landscape,''
  hep-th/0410201 and
M.~R.~Douglas,
  %``The String landscape and low energy supersymmetry,''
%  doi:10.1142/9789814412551_0012
  arXiv:1204.6626 [hep-th].
%
\bibitem{weinberg2} S.~Weinberg,
  %``Anthropic Bound on the Cosmological Constant,''
  Phys.\ Rev.\ Lett.\  {\bf 59} (1987) 2607.
%
\bibitem{Lambda} S.~Perlmutter {\it et al.} [Supernova Cosmology Project Collaboration],
  %``Measurements of Omega and Lambda from 42 high redshift supernovae,''
  Astrophys.\ J.\  {\bf 517} (1999) 565;
A.~G.~Riess {\it et al.} [Supernova Search Team Collaboration],
  %``Observational evidence from supernovae for an accelerating universe and a cosmological constant,''
  Astron.\ J.\  {\bf 116} (1998) 1009.
%
\bibitem{harnik} R.~Kitano and Y.~Nomura,
  %``Supersymmetry, naturalness, and signatures at the LHC,''                                          
  Phys.\ Rev.\ D {\bf 73}, 095004 (2006).
%
\bibitem{comp3} H.~Baer, V.~Barger and D.~Mickelson,
  %``How conventional measures overestimate electroweak fine-tuning in supersymmetric theory,''        
  Phys.\ Rev.\ D {\bf 88}, 095013 (2013).
%
\bibitem{seige} H.~Baer, V.~Barger, D.~Mickelson and M.~Padeffke-Kirkland,
  %``SUSY models under siege: LHC constraints and electroweak fine-tuning,''                           
  Phys.\ Rev.\ D {\bf 89}, 115019 (2014).
%
\bibitem{eenz} J.~R.~Ellis, K.~Enqvist, D.~V.~Nanopoulos and F.~Zwirner,
  %``Observables in Low-Energy Superstring Models,''                                                   
  Mod.\ Phys.\ Lett.\ A {\bf 1}, 57 (1986);
R.~Barbieri and G.~F.~Giudice,
  %``Upper Bounds on Supersymmetric Particle Masses,''                                                 
  Nucl.\ Phys.\ B {\bf 306}, 63 (1988).
%
\bibitem{rns} H.~Baer, V.~Barger, P.~Huang, D.~Mickelson, A.~Mustafayev and X.~Tata,
  %``Radiative natural supersymmetry: Reconciling electroweak fine-tuning and the \                    
%Higgs boson mass,''                                                                                   
  Phys.\ Rev.\ D {\bf 87}, 115028 (2013).
%
\bibitem{ltr} H.~Baer, V.~Barger, P.~Huang, A.~Mustafayev and X.~Tata,
  %``Radiative natural SUSY with a 125~GeV Higgs boson,''                                              
Phys. Rev. Lett. {\bf 109}, 161802 (2012).
%
\bibitem{Chan:1997bi} K.~L.~Chan, U.~Chattopadhyay and P.~Nath,
  %``Naturalness, weak scale supersymmetry and the prospect for the observation o\                     
%f supersymmetry at the Tevatron and at the CERN LHC,''                                                
  Phys.\ Rev.\ D {\bf 58}, 096004 (1998).
%
\bibitem{Barbieri:2009ew} R.~Barbieri and D.~Pappadopulo,
  %``S-particles at their naturalness limits,''                                                       
  JHEP {\bf 0910}, 061 (2009).
%
\bibitem{hgsno} H.~Baer, V.~Barger and P.~Huang,
  %``Hidden SUSY at the LHC: the light higgsino-world scenario and the role of a \                    
%lepton collider,''                                                                                   
  JHEP {\bf 1111}, 031 (2011).
%
\bibitem{upper} H.~Baer, V.~Barger and M.~Savoy,
  %``Upper bounds on sparticle masses from naturalness or how to disprove weak scale supersymmetry,''
  Phys.\ Rev.\ D {\bf 93} (2016) 3,  035016.
%
\bibitem{guts} H.~Baer, V.~Barger and M.~Savoy, arXiv:1602.06973.
%
\bibitem{nuhm2} D.~Matalliotakis and H.~P.~Nilles,
  %``Implications of nonuniversality of soft terms in supersymmetric grand unified\                   
% theories,''                                                                                         
  Nucl.\ Phys.\ B {\bf 435}, 115 (1995);
P.~Nath and R.~L.~Arnowitt,
  %``Nonuniversal soft SUSY breaking and dark matter,''                                               
  Phys.\ Rev.\ D {\bf 56}, 2820 (1997);
J. Ellis, K. Olive and Y. Santoso, Phys.\ Lett.\ B {\bf 539}, 107 (2002);
J. Ellis, T. Falk, K. Olive and Y. Santoso,
Nucl.\ Phys.\ B {\bf 652}, 259 (2003);
H.~Baer, A.~Mustafayev, S.~Profumo, A.~Belyaev and X. Tata,
JHEP {\bf 0507}, 065 (2005).
%
\bibitem{GR} G.~F.~Giudice and R.~Rattazzi,
  %``Living Dangerously with Low-Energy Supersymmetry,''
  Nucl.\ Phys.\ B {\bf 757} (2006) 19.
%
\bibitem{NP} Y.~Nomura and D.~Poland,
  %``Predictive Supersymmetry from Criticality,''
  Phys.\ Lett.\ B {\bf 648} (2007) 213.
%
\bibitem{DM} B.~Dutta and Y.~Mimura,
  %``Landscape of Little Hierarchy,''
  Phys.\ Lett.\ B {\bf 648} (2007) 357.
%
\bibitem{GM} G.~F.~Giudice and A.~Masiero,
  %``A Natural Solution to the mu Problem in Supergravity Theories,''
  Phys.\ Lett.\ B {\bf 206} (1988) 480.
%
\bibitem{KN} J.~E.~Kim and H.~P.~Nilles,
  %``The mu Problem and the Strong CP Problem,''                                                      
  Phys.\ Lett.\ B {\bf 138}, 150 (1984).
%
\bibitem{msy} H.~Murayama, H.~Suzuki and T.~Yanagida,
  %``Radiative breaking of Peccei-Quinn symmetry at the intermediate mass scale,'\                    
%'                                                                                                    
  Phys.\ Lett.\ B {\bf 291}, 418 (1992);
K.~Choi, E.~J.~Chun and J.~E.~Kim,
  %``Cosmological implications of radiatively generated axion scale,''                                
  Phys.\ Lett.\ B {\bf 403}, 209 (1997).
%
\bibitem{radpq} K.~J.~Bae, H.~Baer and H.~Serce,
  %``Natural little hierarchy for SUSY from radiative breaking of the Peccei-Quinn symmetry,''        
Phys.\ Rev.\ D {\bf 91}, 015003 (2015).
%
\bibitem{quevedo} L.~Aparicio, M.~Cicoli, S.~Krippendorf, A.~Maharana, F.~Muia and F.~Quevedo,
  %``Sequestered de Sitter String Scenarios: Soft-terms,''
  JHEP {\bf 1411} (2014) 071.
%
\bibitem{natsumi} K.~J.~Bae, H.~Baer, N.~Nagata and H.~Serce,
  %``Prospects for Higgs coupling measurements in SUSY with radiatively-driven naturalness,''
  Phys.\ Rev.\ D {\bf 92} (2015) 3,  035006.
%
\bibitem{sugra} For a review, see {\it e.g.} 
R.~Arnowitt and P.~Nath,
  %``Developments in Supergravity Unified Models,''
  In *Kane, G.L. (ed.): Perspectives on supersymmetry II* 222-243
  [arXiv:0912.2273 [hep-ph]]  and references therein;
G.~L.~Kane, C.~F.~Kolda, L.~Roszkowski and J.~D.~Wells,
  %``Study of constrained minimal supersymmetry,''
  Phys.\ Rev.\ D {\bf 49} (1994) 6173.
%
\bibitem{mhiggs} M.~S.~Carena and H.~E.~Haber,
  %``Higgs boson theory and phenomenology,''
  Prog.\ Part.\ Nucl.\ Phys.\  {\bf 50} (2003) 63
  [hep-ph/0208209].
%
\bibitem{h125} H.~Baer, V.~Barger and A.~Mustafayev,
  %``Implications of a 125 GeV Higgs scalar for LHC SUSY and neutralino dark matter searches,''        
  Phys.\ Rev.\ D {\bf 85}, 075010 (2012).
%
\bibitem{sw} S.~K.~Soni and H.~A.~Weldon,
  %``Analysis of the Supersymmetry Breaking Induced by N=1 Supergravity Theories,''                   
  Phys.\ Lett.\ B {\bf 126} (1983) 215;
V.~S.~Kaplunovsky and J.~Louis,
  %``Model independent analysis of soft terms in effective supergravity and in string theory,''       
  Phys.\ Lett.\ B {\bf 306} (1993) 269;
A.~Brignole, L.~E.~Ibanez and C.~Munoz,
  %``Towards a theory of soft terms for the supersymmetric Standard Model,''                          
  Nucl.\ Phys.\ B {\bf 422} (1994) 125
   [Erratum-ibid.\ B {\bf 436} (1995) 747];
A.~Brignole, L.~E.~Ibanez and C.~Munoz,
  %``Soft supersymmetry breaking terms from supergravity and superstring models,''                    
  Adv.\ Ser.\ Direct.\ High Energy Phys.\  {\bf 21} (2010) 244
  [hep-ph/9707209].
%
\bibitem{kribs} R.~Harnik, G.~D.~Kribs and G.~Perez,
  %``A Universe without weak interactions,''
  Phys.\ Rev.\ D {\bf 74} (2006) 035006;
C.~J.~Hogan,
  %``Nuclear astrophysics of worlds in the string landscape,''
  Phys.\ Rev.\ D {\bf 74} (2006) 123514;
L.~Clavelli and R.~E.~White, III,
  %``Problems in a weakless universe,''
  hep-ph/0609050.
%
\bibitem{dine} M.~Dine, A.~Kagan and S.~Samuel,
  %``Naturalness in Supersymmetry, or Raising the Supersymmetry Breaking Scale,''
  Phys.\ Lett.\ B {\bf 243} (1990) 250.
%
\bibitem{andre} H.~Baer, V.~Barger, A.~Lessa and X.~Tata,
  %``Discovery potential for SUSY at a high luminosity upgrade of LHC14,''
  Phys.\ Rev.\ D {\bf 86} (2012) 117701.
%
\bibitem{wimp} H.~Baer, V.~Barger and D.~Mickelson,
  %``Direct and indirect detection of higgsino-like WIMPs: concluding the story of electroweak naturalness,''
  Phys.\ Lett.\ B {\bf 726} (2013) 330;
K.~J.~Bae, H.~Baer, V.~Barger, M.~R.~Savoy and H.~Serce,
  %``Supersymmetry with radiatively-driven naturalness: implications for WIMP and axion searches,''
  Symmetry {\bf 7} (2015) 2,  788.
%
\end{thebibliography}
\end{document}